\documentclass[a4paper,fleqn,usenatbib]{mnras}

\usepackage{newtxtext,newtxmath}

\usepackage[T1]{fontenc}
\usepackage{ae,aecompl}

\usepackage{graphicx}	
\usepackage{amsmath}	
\usepackage{amssymb}	
\usepackage{threeparttable}

\title[Bayesian ages in the solar neighbourhood]{Stellar ages and masses in the solar neighbourhood: Bayesian analysis using spectroscopy and Gaia DR1 parallaxes}

\author[Lin et al.]{
Jane Lin,$^{1}$\thanks{E-mail: u5027368@anu.edu.au}
Aaron Dotter,$^{2}$
Yuan-Sen Ting$^{1,3,4,5}$
and Martin Asplund$^{1}$
\\
$^{1}$Research School of Astronomy and Astrophysics, The Australian National University, Canberra, ACT 2611, Australia\\
$^{2}$Harvard-Smithsonian Center for Astrophysics, Cambridge, MA 02138, USA\\
$^{3}$Institute for Advanced Study, Princeton, NJ 08540, USA\\
$^{4}$Department of Astrophysical Sciences, Princeton University, Princeton, NJ 08544, USA\\
$^{5}$Observatories of the Carnegie Institution of Washington, 813 Santa Barbara Street, Pasadena, CA 91101, USA}

\date{Accepted XXX. Received YYY; in original form ZZZ}

\pubyear{2016}

\begin{document}
\label{firstpage}
\pagerange{\pageref{firstpage}--\pageref{lastpage}}
\maketitle

\begin{abstract}
We present a Bayesian implementation of isochrone fitting in deriving stellar ages and masses, incorporating absolute K magnitude ($\rm M_K$) derived from 2MASS photometry and Gaia DR1 parallax and differentiation between initial bulk metallicity and present day surface metallicity, with allowance for incorporating further constraints (e.g., asteroseismology) when available. As a test, we re-computed stellar ages and masses of $\sim 4000$ stars in the solar neighbourhood from six well-studied literature samples using both Hipparcos and TGAS parallaxes. Our ages are found to be compatible with literature values but with reduced uncertainties in general. The inclusion of parallax-based $\rm M_K$ serves as an additional constraint on the derived quantities, especially when systematic errors in stellar parameters are underestimated. We reconstructed the age-metallicity relationship in the solar neighbourhood by re-analysing the Geneva-Copenhagen Survey with the inclusion of TGAS-parallaxes and initial bulk metallicity sampling. We found a flat trend for disk stars with ages $<$11\,Gyr but with smaller scatter at all ages compared to literature. 
\end{abstract}




\section{Introduction}

The field of Galactic archaeology is currently undergoing a revolution thanks in a large part to a new generation of hugely ambitious high-resolution 
spectroscopic Milky Way surveys of $>10^5$ stars, such as 
APOGEE \citep{2016AN....337..863M}, Gaia-ESO \citep{2012Msngr.147...25G} and GALAH \citep{2015MNRAS.449.2604D}, which build on the successes of for example the SEGUE \citep{2009AJ....137.4377Y} and RAVE \citep{2006AJ....132.1645S} surveys at lower spectral resolution. In the near future, the chemical information afforded by such surveys will be complemented by 
exquisite astrometry from the Gaia satellite \citep{2001A&A...369..339P} for even larger stellar samples, which will pinpoint the exact locations in the Galaxy as well as the space motions of the stars. 
Corresponding impressive advances are being made in the modelling of the formation and evolution of galaxies like our own, including a better understanding of star formation and other feedback mechanisms \citep[e.g.][]{2014MNRAS.444.1518V,mcalpine2016eagle}. Unravelling the dynamic, assembly, star formation and chemical history of the Milky Way is within reach using this goldmine of high-quality data and realistic computer simulations \citep[e.g.][]{2002ARA&A..40..487F,2016ApJ...818..130B}. Having accurate age estimates for these stars is naturally highly desirable in this context. 

The determination of accurate stellar ages and masses is notoriously difficult. Various methods exist for determining stellar ages and masses with different levels of applicability and accuracy throughout the Hertzsprung-Russell (HR) diagram \citep[see review by][and references therein]{2010ARA&A..48..581S}. The most widely used and arguably most established method is stellar evolutionary model fitting, where the location of the star on the HR-diagram is compared to age-dependent isochrones/evolutionary tracks. This method is most suited for main sequence turn-off stars and subgiants as isochrones of different ages are well separated in this part of the HR-diagram, making finding the best fitting isochrone relatively straightforward. On the other hand, isochrone fitting falls short in the giant branch and the lower main sequence due to the reduced age sensitivity and degeneracies between stellar parameters. Fortunately, ages of red giants are almost entirely dependent on mass, which can be accurately estimated using asteroseismic frequencies $\rm \Delta\nu$ and $\rm \nu_{max}$ via scaling relations \citep[e.g.,][]{2013ARA&A..51..353C}, or spectroscopic information such as C/N or H lines \citep[e.g.,][]{2016ApJ...823..114N,2016arXiv161003286M,2016A&A...594A.120B}. Determining ages for main sequence stars in general is difficult although some methods exist, especially for young stars by means of chromospheric activity \citep[e.g.,][]{1972ApJ...171..565S} and rotation \citep[e.g.,][]{2014ApJ...780..159E}. 

In this paper, we present \textit{Elli}\footnote{Available at https://github.com/dotbot2000/elli}: a Bayesian Monte Carlo based isochrone fitting code, especially well suited for main sequence turn-off and subgiant stars. \textit{Elli}\footnote{In Norse mythology, Elli is the personification of old age who defeats the god Thor.} aims to provide improved stellar age determinations and error probabilities through the use of priors and by employing the full set of observational constraints together with their uncertainties within a Bayesian framework \citep[e.g.,][]{2004MNRAS.351..487P,2005A&A...436..127J,2013MNRAS.429.3645S,2015A&A...575A..36M,2009AnApS...3..117V,2016arXiv161008071S,2014A&A...570A..66S,2015A&A...577A..90M}. 

The main motivation behind  \textit{Elli} is the upcoming data releases of large wide-field surveys such as GALAH \citep{2015MNRAS.449.2604D}. Combining various sources of information such as spectroscopic and photometric stellar parameters, astrometric data from Gaia  and asteroseismic frequencies from Kepler/K2 \citep{2010PASP..122..131G,2015ApJ...809L...3S} and future missions like TESS \citep{2015JATIS...1a4003R} and PLATO \citep{2014ExA....38..249R}, will undoubtedly increase the accuracy of stellar ages and masses. In particular, the recent Tycho-Gaia Astrometric Solution (TGAS) data release \citep{2016arXiv160904172G} provides accurate parallaxes for two million stars in the Tycho-2 catalogue \citep{2000A&A...355L..27H}. Both the quality and quantity of this data are unprecedented. As a first step, we employ TGAS information as an additional constraint in \textit{Elli} and examine the its effect on the derived ages and masses on several well used and referenced stellar studies.

This paper is structured as follows: Section~\ref{sec:method} describes our implementation of the Bayesian framework. Section~\ref{sec:sample} describes the various literature samples used in this analysis with the results for these stars presented in Section~\ref{sec:results}, including a detailed comparison with literature estimates of the stellar ages and masses and a reanalysis of the age-metallicity relation in the solar vicinity. Finally the conclusions are in Section~\ref{sec:conc}.


\section{Method}\label{sec:method}

\subsection{Bayesian analysis}

In the Bayesian framework, the posterior probability (the probability of parameters given the observed data, $\rm p_1$) is given by the product of the likelihood ($\mathcal{L}$) and prior probability ($\pi$). In our case, the model parameters we want to constrain are age ($\rm \tau$), mass ($\rm M$), initial bulk metallicity ($\rm{[Fe/H]_{bulk}}$), parallax ($\rm \bar{\omega}_{sample}$) and K-band extinction ($\rm A_K$, in our sample of the solar neighbourhood we chose K band as it is least affected by extinction). The model prediction from these five parameters are $\rm T_{eff}$, $\log g$, surface metallicity ($\rm{[Fe/H]_{surf}}$), apparent k magnitude ($\rm m_K$) and parallax   ($\rm \bar{\omega}_{model}$). We will compare these predictions to their observed counterparts. In practice, $\rm \bar{\omega}_{model}$ and the sampled parallax ($\rm \bar{\omega}_{sample}$) are the same, but we chose to make this distinction so that our notations are consistent.

Here we also distinguish between the initial composition of the stellar models, $\rm{[Fe/H]_{bulk}}$, and the $\rm [Fe/H]_{surf}$. We consider the former to be an input parameter and the latter to be a prediction: one which is time-dependent due to the effects of mixing, atomic diffusion and gravitational settling during the evolution of each model star \citep{1994ApJ...421..828T}. The extent to which these effects are accounted for varies from one set of models to another but we argue that it is important to make this distinction. 

Using the information provided in the literature sample, the posterior can be written as:
\begin{multline}
    \   p_1(\tau,M,\rm{[Fe/H]_{bulk},A_K, \bar{\omega}_{sample}},\theta |T_{\rm eff}, {\rm log}g, {\rm [Fe/H]_{surf}}, {\rm m_{K}}, \bar{\omega}_{model},\mathcal{D}) \\  \propto \mathcal{L}(\tau,M,\rm{[Fe/H]_{bulk}}, \rm A_K, \bar{\omega}_{sample},\theta)  \pi
	\label{eq:ext}
\end{multline}

Where $\pi$ is the prior, $\mathcal{L}$ is the likelihood and $\mathcal{D}$ is any additional measurements of the stellar properties which can be used to constrain its parameters (e.g., asteroseismic parameters), likewise $\rm \theta$ is any additional model parameters we might want to sample in the future. If the uncertainties of the observed parameters are Gaussian, the likelihood $\mathcal{L}$ is defined as:

\begin{equation}
\mathcal{L}=\frac{1}{\sqrt{(2\pi)^n \mathrm{det}(\Sigma)}} {\rm exp}\Big( -\frac{1}{2}(\vec{O}-\vec{S})^T \Sigma^{-1} (\vec{O}-\vec{S})  \Big)
	\label{eq:mat}
\end{equation}
Here $\vec{O}$ is a vector consisting of the $n$ observed parameters ($\{\rm T_{\rm eff}, \log g, [Fe/H]_{surf}, m_K, \bar{\omega}_{Gaia}, \mathcal{D}\}$), $\vec{S}$ is the vector of the corresponding model predictions and $\Sigma$ is the covariance matrix of the observed parameters. Here we assume the reported parameters have no covariances, Equation~\ref{eq:mat} reduces to the familiar form:

\begin{equation}
    \mathcal{L}=\prod_{i} \frac{1}{\sqrt{2 \pi} \sigma_i} \times {\rm exp} \Big(-\frac{ (O_i -S_{i})^2 }{2 \sigma_{i}^2} \Big)
	\label{eq:lik}
\end{equation}

Ideally all observational data which help constrain $\rm p_1$ and override any unrealistic parameters should be considered. An example of this is the age dependence on erroneous $\log g$ is reduced when $\rm m_K$ is taken into account (see Section~\ref{sec:adi}) or  by the inclusion of asteroseismic $\rm \nu_{\rm max}$ and $\rm \Delta \nu$. Likewise, it is possible to include different values, for example $\rm T_{eff}$ obtained by different methods, each with their own uncertainties. 

For this work, we analyse mostly literature spectroscopic $\rm T_{eff}$, $\log g$, and [Fe/H] as well as Hipparcos/TGAS parallaxes and 2MASS photometry along with their respective uncertainties; these come with no covariance terms and so we are left with a diagonal covariance matrix for the tests carried out herein. But we emphasise that \textit{Elli} has the ability to handle non-diagonal covariance  if necessary. Indeed, it would be ideal to incorporate full covariance into such analyses, as demonstrated in \citet{2016arXiv161008071S}.

For the prior probability in mass, we adopt a flat initial mass function (this mass range is chosen to span our isochrone grid):
	
\[   
\pi(M) = 
     \begin{cases}
       \text{1} &\quad\text{for}\ 300 M_{\odot} > M > 0.1 M_{\odot}\\
       \text{0} &\quad\text{else} \\
     \end{cases}
\]

We have also explored the possibility of including the Salpeter IMF and found it has little impact on the resulting ages. For parallax ($\rm \bar{\omega}_{sample}$), we simply require it to be positive. We allow extinction ($\rm A_K$) to vary from 0 to a maximum, which is provided by the 2D dust map presented in \citep{1998ApJ...500..525S}. We chose not to include any priors for star formation rate and metallicity as we do not have universally established relationships for them. Furthermore, as discussed in \citet{2005A&A...436..127J}, the quality of the data determines the influence of priors: high quality observations mean the derived values only have a weak dependence on priors, whereas low quality observations have much more prior-dependent outcomes. Hence we only adopt the most conservative assumptions as we are ultimately trying to use our analysis to gain further insight into the metallicity distribution function, star formation history, and the initial mass function of the solar neighbourhood. However our code is flexible enough such that additional priors can be easily added.

\subsection{Stellar isochrones}

We use a grid of isochrones from the MESA Isochrones and Stellar Tracks (MIST) project \citep{2016ApJ...823..102C}, with $-4 \le \rm{[Fe/H]_{bulk}} \le +0.5$ and $\rm [\alpha/Fe] = 0$ (we will be implementing $\rm [\alpha/Fe] \neq 0$ isochrones as soon as they are available). The grid has an age range of log(age\,[yr]) from 5 to 10.3 and an initial mass range of 0.1 to 300\,$\rm M_\odot$. These models adopt the protosolar abundances of \citet{2009ARA&A..47..481A}. Diffusion and gravitational settling are calculated in the formalism of \citet{1994ApJ...421..828T}. Low temperature ($\rm log(T_{eff})<4$) opacities  are taken from \citet{2005ApJ...623..585F} and high temperature ($\rm log(T_{eff})>4$) opacities are taken from \citet{1996ApJ...464..943I}. The equations of state are taken from \citet{2002ApJ...576.1064R}, \citet{1995ApJS...99..713S} and \citet{2000ApJS..126..501T}. The surface boundary condition is computed using ATLAS12 models \citep{1993sssp.book.....K}, covering $\rm T_{eff}$ from 2500 to 50000\,K and $\log g$ from 0 to 5\,dex. In addition, \textit{Elli} also has Dartmouth Stellar evolution database isochrones \citep{2008ApJS..178...89D} implemented.

\subsection{Markov chain Monte Carlo}\label{sec:mcmc}

Markov chain Monte Carlo (MCMC) is used to sample the posterior via the Python package \textit{Emcee} \citep[][]{2013PASP..125..306F}.  \textit{Emcee} implements the affine-invariant MCMC ensemble sampler wherein an ensemble of random walkers sample the parameter space; the walkers are initially distributed as a cloud surrounding some initial point in parameter space and the next step for each individual walker is determined by looking at the progress of the ensemble \citep{goodman}. The ensemble approach was designed to sample a large parameter space more efficiently than could be done by single walker over the same total number of steps. In this work we deploy 200 walkers for each star, constructing chains of 1000 steps per walker, sampling the posterior a total  of 200,000 times. An initial burn-in phase of 200 steps is built into each chain. Our experiments have shown this number of steps is enough to achieve convergence for most of our samples.

In each step of the MCMC, we will sample model age, mass and bulk metallicity, parallax and extinction. These five parameters will produce model predictions in $\rm  T_{\rm eff}$, $ {\rm log}g$, $\rm [Fe/H]_{surf}$,  $\rm m_K$ and $\rm \bar{\omega}_{model}$. $\rm \bar{\omega}_{model}$ will be compared to $\rm \bar{\omega}_{Gaia}$ and we use the magnitude formula with sampled extinction and parallax to convert the model $\rm M_K$ to apparent magnitude for comparison. The advantage of using MCMC is that we are able to sample the full distributions of ages and masses as well as the model predictions  ($\rm S_i$) for every star. Hence it is possible to calculate full statistics for all involved parameters. 

The initial guess in age and mass is calculated by firstly finding the closest two sets of isochrones encompassing the measured metallicity. For an isochrone set, we calculate the logarithm of the likelihood following Equation~\ref{eq:lik}, given the observed values (assuming Gaussian uncertainties). Age and mass are the weighted averages of ages and masses of each isochrone in the set, with the weights being the likelihood. The final age and mass guess is the interpolated values at the given metallicity. If no suitable age/mass guesses are found (for example when the star has peculiar stellar parameters, the likelihoods are very small), the algorithm returns 5\,Gyr for age and 1\,$\rm M_\odot$ for mass.

The convergence of MCMC chains can be gauged by three outputs: the acceptance fraction, the autocorrelation time and the Gelman-Rubin scale reduction factor. The acceptance fraction is the fraction of proposed steps that are accepted \citep{2013PASP..125..306F}, essentially a measure of the validity of the chain. \citet{2013PASP..125..306F} indicate that a general target value is between 0.2 and 0.5 while a value close to either 0 or 1 is unreliable. We found our acceptance fractions are generally between 0.2 and 0.6, with 0.1 being the threshold for a robust chain. The autocorrelation time is an estimate of the number of steps required by the walkers to draw independent samples from the posterior. \citet{2013PASP..125..306F} suggests that the samplers should be run for a few autocorrelation times. Finally the \citet{10.2307/2246093} scale reduction factor is a measure of the similarity between chains, with the assumption that at convergence, all MCMC chains will be very similar and the factor approaches 1. The convergence of our samples is discussed in Section~\ref{sec:results}.

One potential limitation of our MCMC implementation is the sampling of multimodal distributions: if the posterior has modes separated by large valleys of low probabilities, it is possible for the walkers to get stuck near one mode. To combat this problem, we deliberately force the initial clouds to be large ($\pm$40\% of the initial guesses), such that atleast some walkers should escape the dominant mode. To see if multimodality is an actual issue for our sample, we picked 100 stars located on the various parts of the HR-diagram of our sample and ran each star with 50 different initialisations (with initial ages and masses ranging 0.1-20\,Gyr and 0.3-2\,$\rm M_\odot$). We found that most of the initialisations did converge to a consistent age and mass, with very little dispersion between the results of the different initialisations. The only initial conditions which did not converge are physically impossible (e.g., age of 15\,Gyr and 2\,$\rm M_\odot$). In addition to internal tests, we find our results are comparable to literature samples which do not use MCMC and therefore do not suffer from MCMC multimodality issues. We thus conclude that at least for the sample at hand, our MCMC implementation can sample the posterior effectively. However we do acknowledge that \textit{Emcee} is not a global method and we recommend always to examine the MCMC distributions of stars which have problematic convergences, as indicated by the aforementioned metrics.

\section{Observational data}\label{sec:sample}

\begin{figure}
	\includegraphics[width=\columnwidth]{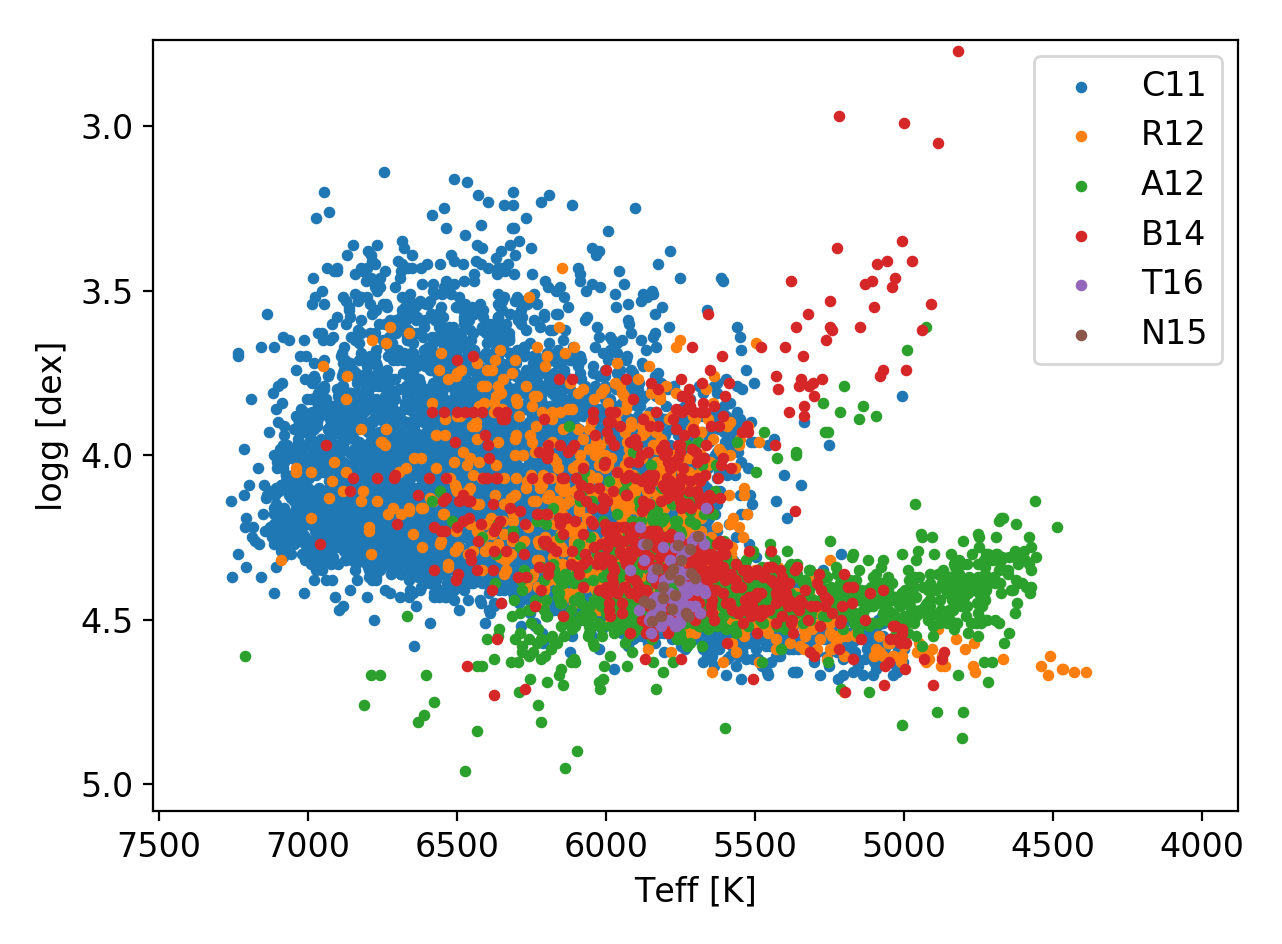}
    \caption{HR-diagram of our combined sample. The upturn in the lower main sequence of the A12 sample signals erroneous $\log g$ values.}
    \label{fig:hr}
\end{figure}

To test our methodology, we use the samples of \citet{adi} (hereafter A12), \citet{ben} (hereafter B14), \citet{ram} (hereafter R12), \citet{nis} (hereafter N15) and \citet{2016A&A...590A..32T} (hereafter T16), all with spectroscopically determined stellar parameters (Figure~\ref{fig:hr}). Furthermore, we obtain a set of stars from the Geneva-Copenhagen Survey \citep[][]{2004A&A...418..989N}, re-analysed with more accurate temperatures from the infrared flux method and a new metallicity scale \citep[][ hereafter C11]{2011A&A...530A.138C}. Stellar ages and masses are also available for our samples by the original authors with the exception of the A12 set. Literature ages were determined by comparing observed parameters to theoretical stellar evolutionary models; we refer to the respective studies for a discussion on the individual methods and stellar evolution models used. \textit{Elli} has also been used to compute ages for the RAVE sample \citep[][]{2017AJ....153...75K}, see \citet{2018MNRAS.475.1203C}.

The HR-diagram of the A12 sample shows a clear upturn in $\log g$ at the beginning of the lower main sequence, in contrast to monotonic decrease expected from stellar models; this feature has been noticed by others (e.g., B14). The exact reason for this upturn is not well understood but may be related to the breakdown in 1D LTE ionisation equilibrium for K dwarfs  in the field \citep[e.g.,][]{1998A&AS..129..237F,2004A&A...420..183A} and open clusters \citep[e.g.,][]{2006AJ....131.1057S,2010PASP..122..766S}, the reason for which is still unknown. The effect of such erroneous $\log g$ on derived ages is discussed in Section~\ref{sec:adi}.

In addition to stellar parameters, we adopt apparent K magnitudes from 2MASS \citep[][]{2003yCat.2246....0C}, and  parallaxes from Hipparcos \citep[][]{2007A&A...474..653V} and TGAS \citep{2016arXiv160904172G} (when available). The K band is chosen because it is little affected by interstellar reddening and is widely available. 

\section{Results}\label{sec:results}
\begin{figure*}
	\centering
	\includegraphics[width=180mm]{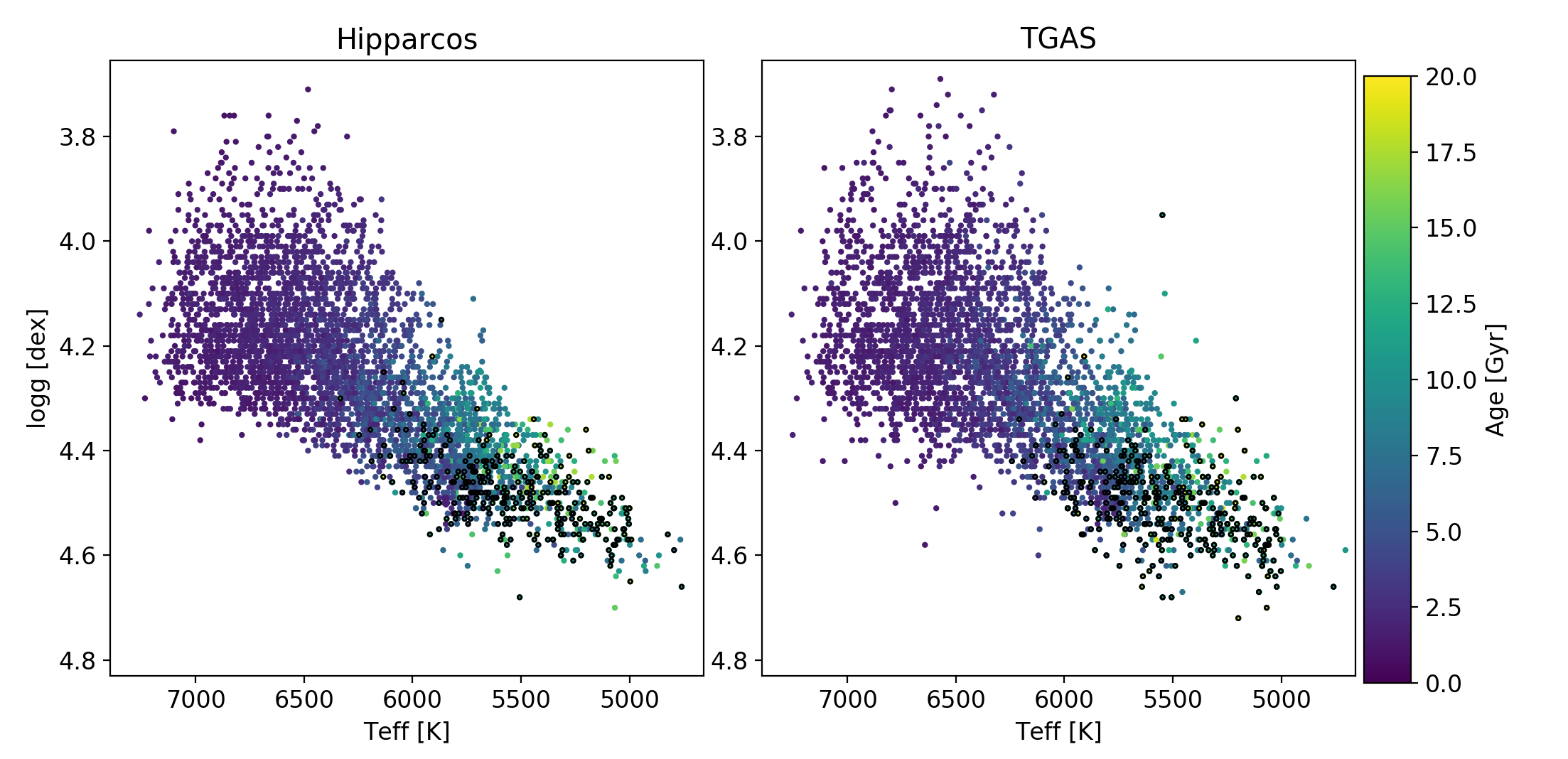}
    \caption{{\bf Left panel:} HR-diagram of the combined literature sample (excluding A12), colour coded in \textit{Elli} derived ages using Hipparcos-based $\rm M_K$. Black circles indicate the 320 stars which the absolute age differences between literature and \textit{Elli} are greater than 3\,Gyr. {\bf Right panel:} Same, but using TGAS parallaxes, with 325 problematic stars overplotted. We have examined every outlying case for both Hipparcos and TGAS parallax ages and are confident that the stars are well fitted.} 
    \label{fig:hr_all}
\end{figure*}

In total, we derived ages and masses for 4602 stars with Hipparcos based parallaxes and 4509 stars with TGAS based parallaxes; 3568 stars have both sets of parallaxes (Figure~\ref{fig:hr_all} shows the H-R diagrams of both samples). Table~\ref{tab:summary} shows the breakdown by literature samples. We treat the common stars between literature samples as separate stars since their stellar parameters differ from sample to sample. 

\begin{figure}
	\includegraphics[width=\columnwidth]{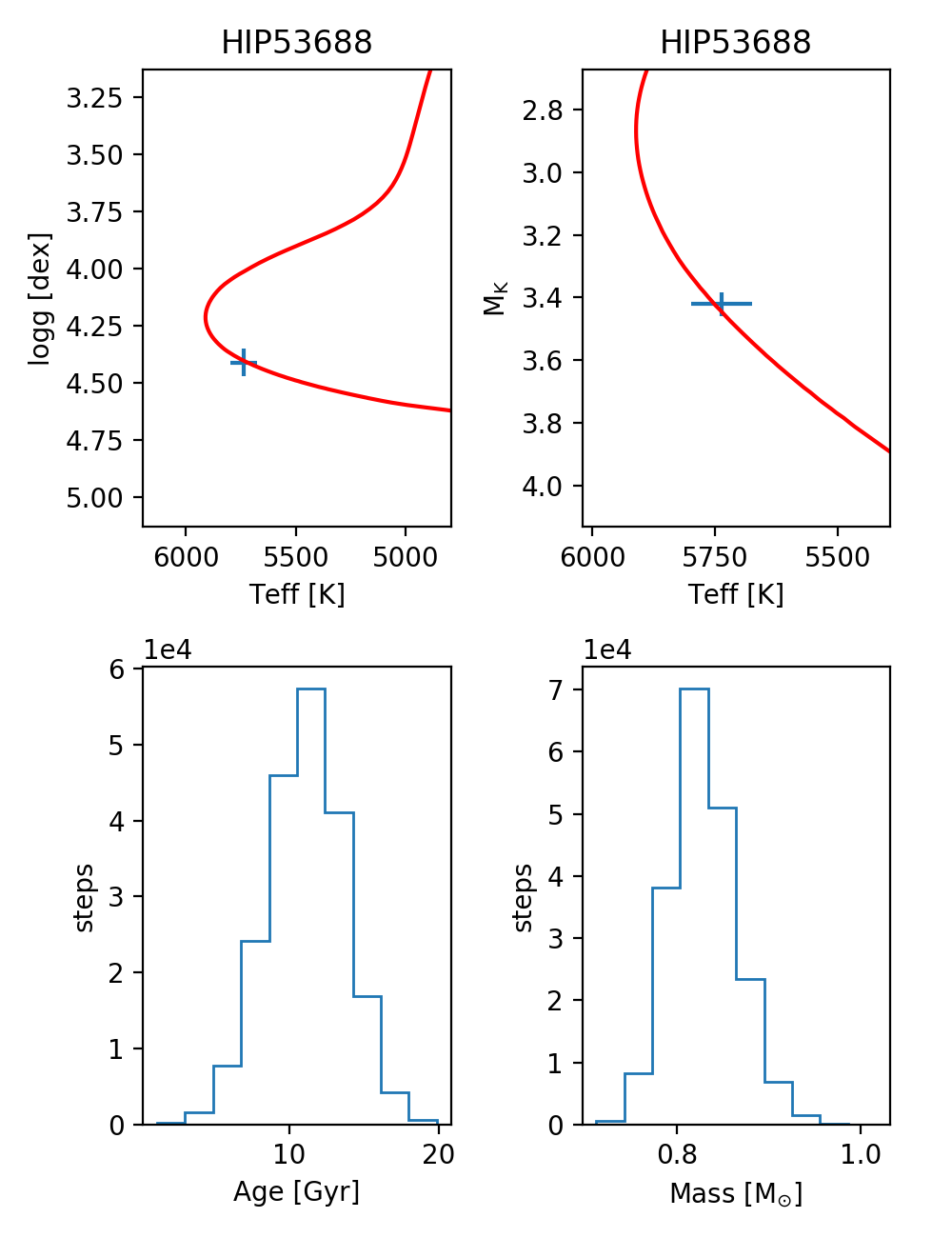}
    \caption{Example \textit{Elli} output for a star. {\bf Top panels:} location of the star (HIP53688) in the  $\rm T_{eff}$-$\log g$ and $\rm T_{eff}$-$\rm M_K$ planes (using TGAS parallaxes), with the best fitting isochrone plotted (11.13\,Gyr) at $\rm [Fe/H]_{bulk}=$-0.32\,[dex].  {\bf Bottom panels:} the posterior distributions for age and mass. The star is well fitted in both planes. See text for details. }
    \label{fig:lnp}
\end{figure}

 Figure~\ref{fig:lnp} shows as an example the \textit{Elli} output for one star (HIP53688). The top panels show the location of the star in the $\rm T_{eff}$-$\log g$  and $\rm T_{eff}$-$\rm M_K$ planes. The bottom panels show posterior distributions as a function of sampled ages and masses (similar distributions are recovered for all sampled and observed parameters). 
The mean of this distribution is taken as the most probable value and the standard deviation as the uncertainty; we note that one advantage with this type of modelling is that the full probability distribution function is available should it be desired to use. Top panels in Figure~\ref{fig:lnp} show excellent agreement between our ages and isochrones in both the $\rm T_{eff}$-$\log g$ and $\rm T_{eff}$-$\rm M_K$ planes. Our age and mass for Hipparcos parallax is 11.18$\pm$2.57\,Gyr, 0.82$\pm$0.03\,$\rm M_\odot$ and for TGAS parallax is 11.13$\pm$2.58\,Gyr, 0.83$\pm$0.04\,$\rm M_\odot$. The corresponding literature value from C11 for this star is $\rm \tau$=11.69$\pm$4.79\,Gyr , $\rm M$=0.89$\pm$0.05\,$\rm M_\odot$, again the agreement is excellent for both parallaxes (22.43$\pm$0.33\,mas and 21.48$\pm$0.69\,mas for TGAS and Hipparcos respectively). 
  
As discussed in Section~\ref{sec:mcmc}, we impose some cuts to our results to maximise convergence. Firstly, we only take walkers which have autocorrelation time lesser than 30\% of the running time \citep[][]{2013PASP..125..306F} and acceptance fraction greater than 0.1. Secondly, we impose a cut of 1.2 \citep{brooks1998markov} on the scale reduction factor in bulk metallicity and parallax. The rational being that the scale reduction factor is a measure of similarities among chains and is calculated for all  sampled parameters. For stars with multimodal distributions, the scale reduction factors will be large in age and mass, but will remain small for bulk metallicity and parallax, as it is less likely to be multimodal. For a selection of stars we also examined the chains individually and found similar results from large segments of chains and different length of chains (i.e. chains yielding similar results for 200 steps vs 1000 steps vs 2000 steps) - both are signs of convergence.

Finally, we impose a relative uncertainty cut of 0.5 in age. Potential multimodal stars are flagged if they have age reduction factors greater than 5 or the number of walkers which have run more than 30\% of the autocorrelation time is below 20 (out of 200). These cuts are found by examining the full MCMC distributions of the R12 sample. Our results for both TGAS and Hipparcos parallaxes are summarised in Table~\ref{tab:summary}. The HR diagram of the TGAS sample colour coded in \textit{Elli} ages is shown in right panel of Figure~\ref{fig:hr_all}.

\begin{table*}
\centering
\begin{minipage}{180mm}
\caption{Summary of comparisons between \textit{Elli} and literature parameters for various samples. Differences are calculated as literature - \textit{Elli} values.}
\label{tab:summary}
\begin{tabular}{lllllll}
       & Hipparcos & &  & TGAS  & &  \\
Sample & 
$\overline{\rm{ \Delta Age}}$, $\rm{\sigma(\Delta Age)}\ \rm{[Gyr]}$ &  $\overline{\rm{ \Delta M}}$, $\rm{\sigma(\Delta M)} \ \rm{[M_\odot]}$ & stars & $ \overline{\rm{\Delta Age}}$, $\rm{ \sigma(\Delta Age)}\ \rm{[Gyr]}$ &  $\overline{ \rm{\Delta M}}$, $\rm{\sigma(\Delta M)}\ \rm{[M_\odot]}$ & stars \\
\hline
\citet{adi}\textsuperscript{a}   & N/A  & N/A & 613   & N/A  & N/A  & 721   \\
\citet{ben}   & -1.08, 1.52  &  0.019, 0.043   & 261   & -1.59, 1.87  &  0.019, 0.033     & 188   \\
\text{\citet{2011A&A...530A.138C}}\textsuperscript{b}    & -0.97, 1.43&  0.034, 0.037 & 3120  & -1.05, 1.48 & 0.030, 0.053  & 3070  \\
\citet{nis}   & -1.04, 0.38& 0.041, 0.008 & 17    & -1.35, 0.42    & 0.039, 0.008   & 14    \\
\citet{ram}\textsuperscript{c}   & -0.94, 1.64   &    0.030, 0.024    & 532  & -1.06, 1.58  &     0.026, 0.027   & 466   \\
\text{\citet{2016A&A...590A..32T}}   & -1.01, 0.66 &  N/A  & 57    & -1.19, 0.66 &    N/A   & 48 \\
\hline 
\end{tabular}
    \begin{tablenotes}
      \small
      \item \textsuperscript{a} $\log g$ is excluded in the analysis for being unrealistic
      \item \textsuperscript{b} Our ages are compared with Padova \citep{2008A&A...484..815B} ages
      \item \textsuperscript{c}  60 stars have no known literature ages
    \end{tablenotes}
\end{minipage}
\end{table*}

\subsection{Comparison with literature values}\label{sec:lit}

\begin{figure*}
	\centering 
	\includegraphics[width=180mm]{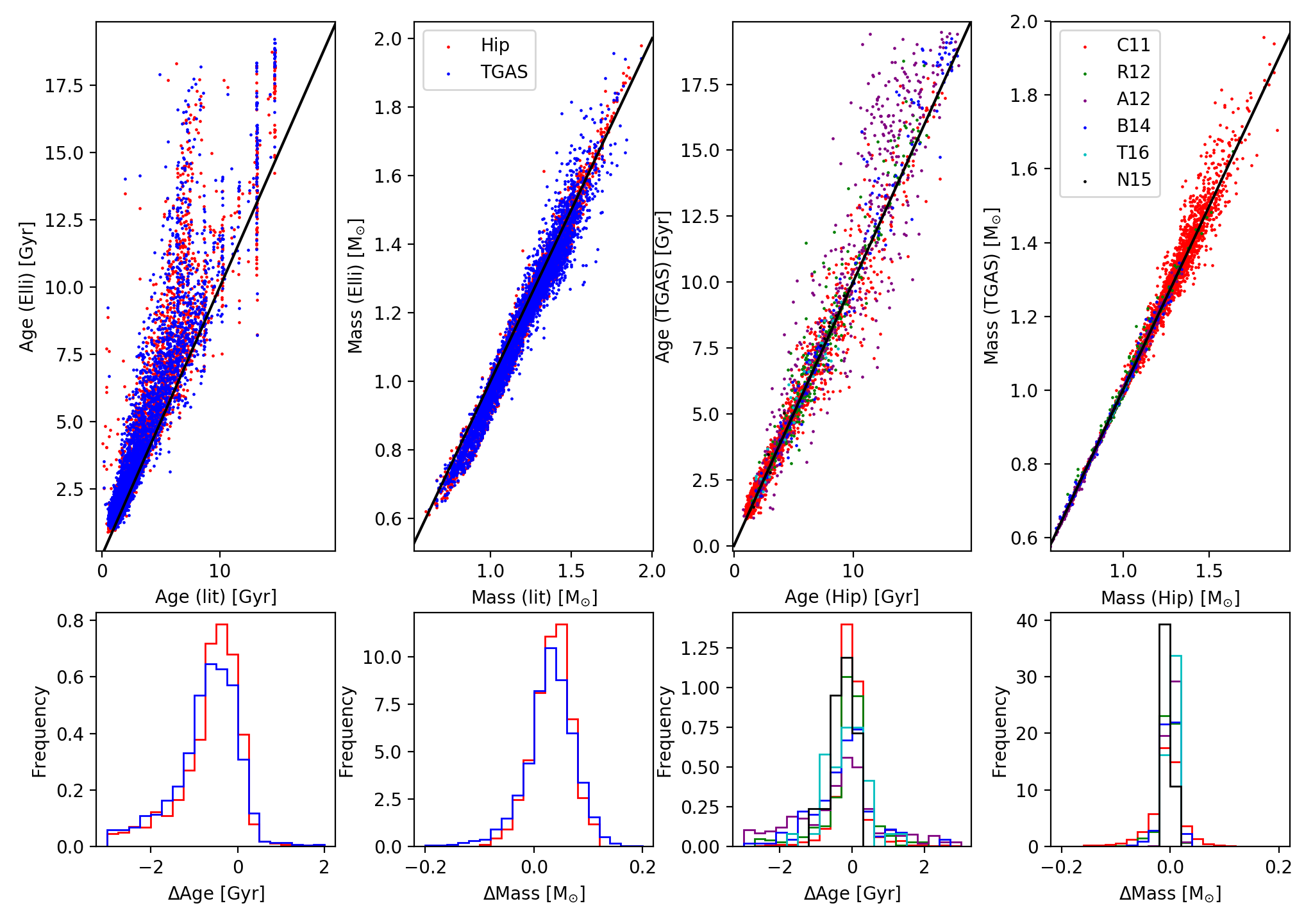}
    \caption{Comparison of between our results with literature values. On average, our ages are older than literature ages due to our choice of isochrones and $\rm [Fe/H]_{bulk}$ sampling. {\bf Top panels:} (from left to right) {\bf (1)} Comparison between the various literature ages (x-axis) and ages derived in this work (y-axis).  {\bf (2)} Same as in (1) but for mass. {\bf (3)} Comparison between ages derived from this work using Hipparcos (x-axis) and TGAS (y-axis) parallaxes. {\bf (4)} Same as in (3) but for mass. {\bf Bottom panels:} (from left to right) {\bf (1)} Histogram of the difference in ages between this work and literature (literature ages$-$\textit{Elli} ages), colours are the same as its top panel. {\bf (2)} same as (1) but for mass. {\bf (3)}  Histogram of the differences in ages derived using Hipparcos and TGAS parallaxes (Hipparcos ages$-$TGAS ages) in this work, colours are the same as its top panel. {\bf (4)} Same as (3) but for mass. The T16 sample does not report masses.} 
    \label{fig:diff}
\end{figure*}

In this Section we present the comparison between our derived stellar ages and masses with literature values for the B14, R12, N15, T16 and C11 samples; the A12 sample is discussed in Section \ref{sec:adi} as they did not provide age/mass estimates themselves. Figure~\ref{fig:diff} shows the comparison between our ages/masses and literature ages/masses for 3568 stars with both Hipparcos and TGAS parallaxes (excluding the A12 sample). The bias and standard deviation in ages and masses are reported in Table~\ref{tab:summary} for the individual samples. 

\begin{figure}
	\includegraphics[width=\columnwidth]{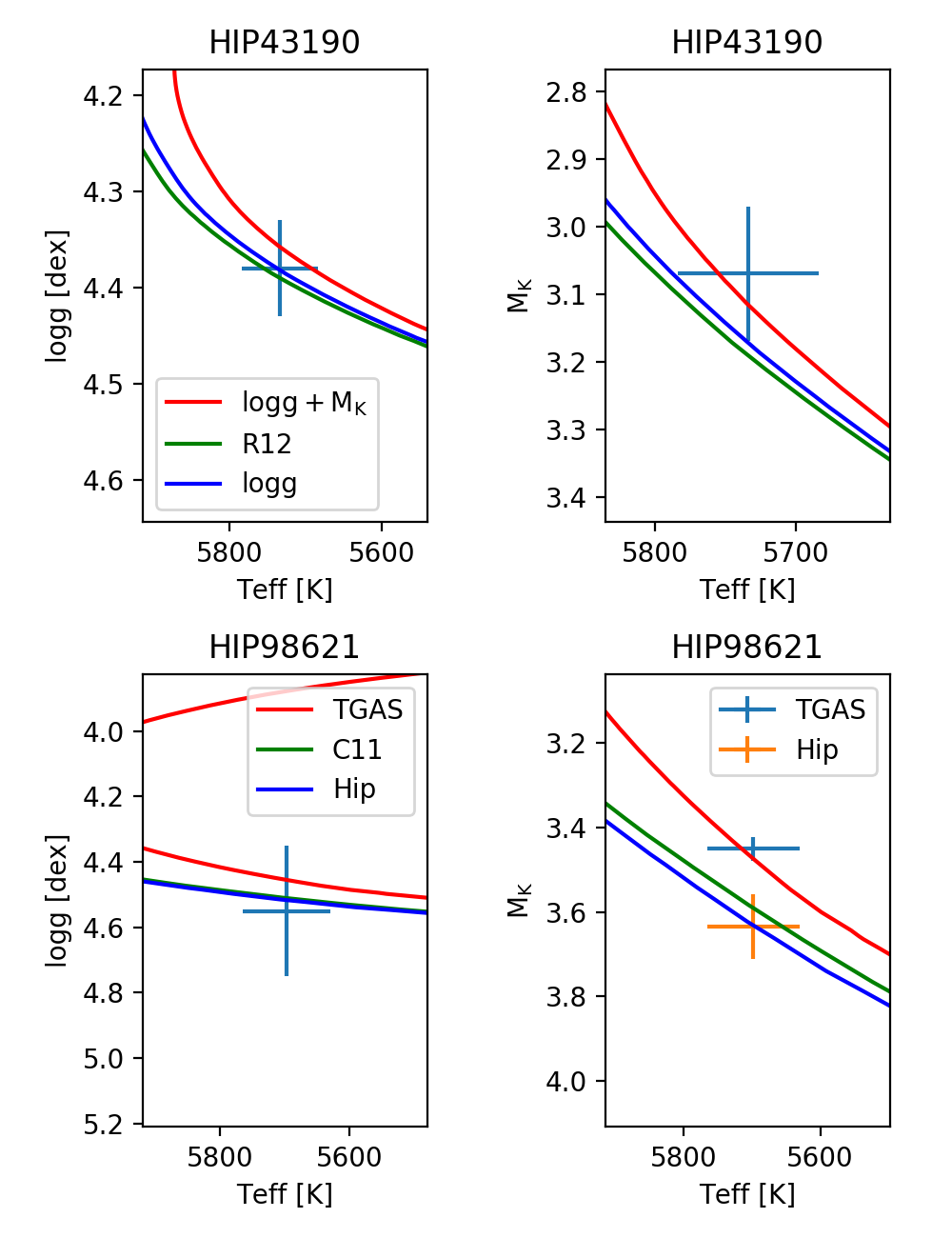}
    \caption{Examples of $\rm M_K$ inducing large age differences when compared to literature (top panels), TGAS  $\rm M_K$ in particular is further skewing TGAS ages away literature values (bottom panels). {\bf Top panels:} left, location of the star HIP43190 in the $\rm T_{eff}$-$\log g$ plane. The isochrones are: red=7.35\,Gyr (best fit with TGAS-based $\rm M_K$ and logg), green=6.29\,Gyr (R12), blue=6.58\,Gyr (best fit with logg only). Right, same, but in the $\rm T_{eff}$-$\rm M_K$ plane, with the green point representing the TGAS-based $\rm M_K$. {\bf Bottom panels:} left, location of the star HIP98621. The isochrones are: red=6.90\,Gyr (best fit with TGAS-based $\rm M_K$), green=3.62\,Gyr (C11), blue=3.96\,Gyr (best fit with Hipparcos-based $\rm M_K$).  Right, same,  the cyan and orange points represent $\rm M_K$ derived from TGAS and Hipparcos parallaxes, respectively. }
    \label{fig:hrs}
\end{figure}

As indicated by Figure~\ref{fig:diff}, overall there is a good agreement between \textit{Elli} and literature values and we do not observe any systematic differences between Hipparcos and TGAS derived ages and masses.  Most the differences between our ages/masses and literature values is due to our inclusion of parallax-based $\rm M_K$ in the analysis, but additional systematic age differences also arise from different choices of isochrones (especially when MIST currently has no alpha-enhancement for metal poor stars)   and our sampling of initial metallicity.  Literature ages are derived without fitting for $\rm [Fe/H]_{bulk}$, thus can be significantly biased \citep{2017arXiv170403465D}. Figure~\ref{fig:hrs} shows the effect of $\rm M_K$ on age for a typical star: our age (6.58\,Gyr, blue line) agrees quite well with R12 age (6.29\,Gyr, green line) when $\rm M_K$ is disabled (top left). Once enabled, our age differs significantly (7.35\,Gyr, red line), as $\rm M_K$ suggests substantially older ages (top right).  

Surprisingly, on average ages and masses calculated using Hipparcos parallaxes agree slightly better with literature than those calculated using TGAS parallaxes. We observe that larger age deviation from literature for the TGAS stars is likely due to the inconsistency between spectroscopic $\log g$ and photometric $\rm M_K$. On average, the uncertainties associated with TGAS parallaxes are smaller compared to Hipparcos parallaxes (hence smaller uncertainties in $\rm M_K$). This means when $\log g$ and $\rm M_K$ do not agree, MCMC favours the parameter which has less uncertainty, hence driving ages away from purely $\log g$-based estimates. On the other hand, the larger uncertainties of Hipparcos parallaxes mean the aforementioned inconsistency has a lesser effect on derived ages. An example of such a case is shown in Figure~\ref{fig:hrs} (bottom panels): both spectroscopic parameters  and Hipparcos-based $\rm M_K$ (orange point, bottom right panel) have large uncertainties, so that our age (3.96\,Gyr, blue) is close to the literature value (3.62\,Gyr, green, C11), albeit they are still different due to $\rm M_K$. When we switch to TGAS-based $\rm M_K$ (cyan point, bottom right panel), two changes are observed: firstly, it shifts $\rm M_K$ up and secondly, the $\rm M_K$ uncertainty is reduced significantly. These changes effectively force the age to deviate even further from literature (our age is now 6.90\,Gyr). 

Finally, as expected, the TGAS parallaxes are associated with smaller relative uncertainties (  $\rm  \overline{\sigma \tau}=$ 0.23\,Gyr and mean $\rm \overline{\sigma  M}=$ 0.033\,$\rm M_\odot$, respectively) compared to Hipparcos (0.32\,Gyr and 0.036\,$\rm M_\odot$, respectively). The mean absolute difference between Hipparcos and TGAS parallaxes is 0.86\,mas, with a standard deviation of 1.15\,mas for our combined sample. The \textit{Elli} statistical uncertainties are substantially smaller compared to that of the literature, which are 0.72\,Gyr in age and 0.065\,$\rm M_\odot$ in mass, respectively. 

In total, there are 320 (for Hipparcos-based $\rm M_K$) and 325 (for TGAS-based $\rm M_K$) stars with absolute age differences greater than 3\,Gyr between our results and literature. They are represented as black circles in Figure~\ref{fig:hr_all}. We examined, on a star-by-star basis, our best fitting isochrones against isochrones of literature ages in both $\rm T_{eff}$-$\log g$ and $\rm T_{eff}$-$\rm M_K$ planes. We found most of these stars have substantial uncertainties in stellar parameters such that both \textit{Elli} and literature ages can be considered reasonable (e.g., lower left panel in Figure~\ref{fig:hrs}). Not surprisingly, most of these stars are located on the main sequence-i.e. areas of low stellar age sensitivity. For the remaining cases,  $\rm M_K$  seems to be the driving factor in the age discrepancy, as discussed previously. Isochrones of \textit{Elli} ages provide consistently better fits in the $\rm T_{eff}$-$\rm M_K$ plane. For the rest of the sample, we examined the goodness of the fit of our ages on both $\rm T_{eff}$-$\log g$ and $\rm T_{eff}$-$\rm M_K$ planes. Most stars are well fitted, with few exceptions being stars with clearly inconsistent spectroscopic $\log g$ and photometric $\rm M_K$.

We note the presence of vertical spikes near 7, 10, 13 and 14.5\,Gyr in the top left-most panel in Figure~\ref{fig:diff}. They are from the B14 sample, indicating a potential grid effect in their analysis. In the same figure, there is also a trail of stars with very low literature ages (< 1\,Gyr) and high \textit{Elli} ages. This consists entirely of stars from the R12 sample, where rotation period was used to derive the ages for very young stars. Plotting best matching isochrones on the HR-diagram indicates our ages are well fitted for this sample. The agreement between our ages and those presented in C11 (the bulk of the literature sample) is especially good. Other than small systematic offsets in age and mass (likely due to the choice of isochrone grids), the average differences for age and mass are very small. The agreement is probably due to the C11 sample being composed of mostly turn-off stars, where isochrone fitting is most effective. The bifurcation observed in top left panel in Figure~\ref{fig:diff} is mainly due to larger \textit{Elli} ages for relatively metal-poor main sequence stars in the C11 sample. We have examined their isochrone fits and found them to be well fitted. These stars represent the high age, metal-poor tail of the AMR in Figure~\ref{fig:amr}.

\subsection{The \citet{adi} sample}\label{sec:adi}

As discussed in Section~\ref{sec:sample}, the lower main sequence with the A12 spectroscopic parameters exhibit a suspicious upturn, implying unrealistic surface  gravities. This section discusses the impact of such erroneous $\log g$ on age. 
 
\begin{figure}
	\includegraphics[width=\columnwidth]{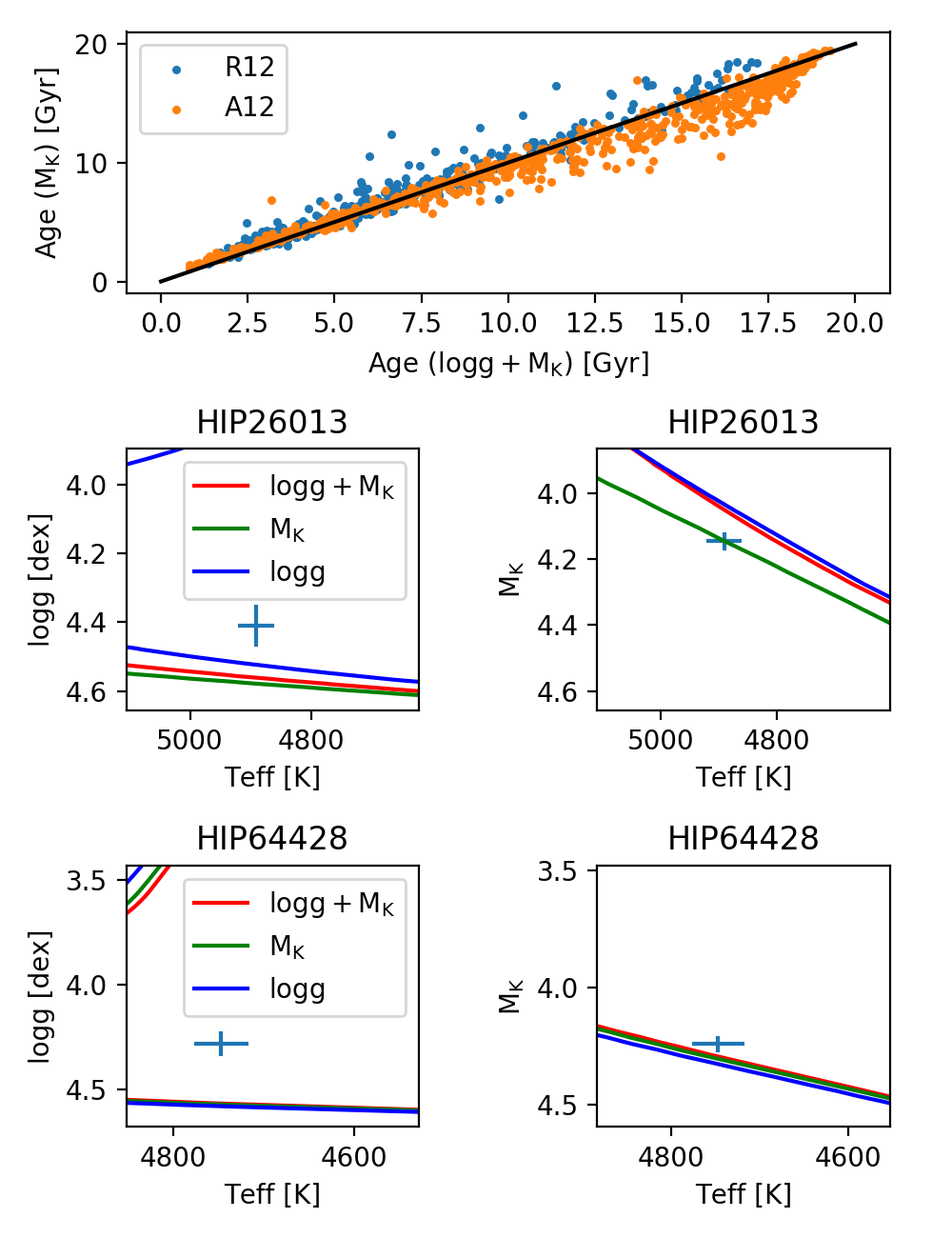}
    \caption{Effect of $\rm M_K$  on age when logg is unrealistic due to the breakdown of 1D LTE ionisation equilibrium in K dwarfs. {\bf Top panel:} Comparison between ages computed using both literature $\log g$ and TGAS-based $\rm M_K$ (x-axis) and ages computed when $\log g$ is excluded (y-axis). Orange points are the A12 sample, blue points are the R12 sample, as control. {\bf Middle left panel:} Location of the star HIP26013 in the $\rm T_{eff}$-$\log g$  plane. Isochrones are: green=7.25\,Gyr (best fit with $\rm M_K$ only), red=10.05\,Gyr (fit with both $\log g$ and $\rm M_K$ included) and blue=15.63\,Gyr (best fit with $\log g$ only). {\bf Middle right panel:} Same as middle left, but in the $\rm T_{eff}$-$\rm M_K$ plane. {\bf Bottom panels:} Same as middle panels for the star HIP64428. The isochrones are blue=16.44\,Gyr, red=18,43\,Gyr, green=17.64\,Gyr.}
    \label{fig:adi}
\end{figure}

Figure~\ref{fig:adi} shows the comparison between \textit{Elli} ages computed using (a) both literature $\log g$ and TGAS-based $\rm M_K$ (x-axis) and (b) ages computed when $\log g$ is excluded (y-axis), for a set of 658 A12 stars (obtained after the same cuts as the rest of the literature samples). As a control, we also performed the same exercise for a set of 398 R12 stars, where spectroscopic $\log g$ values are more consistent with stellar model predictions. We observe that the A12 stars show a clear preference towards older ages when $\log g$ is included in the analysis, whereas the R12 stars show no such trends (the few stars with high ages without $\log g$ are due to $\rm M_K$ forcing higher ages). This preference for older ages is due to the underestimated $\log g$. The red and blue isochrones in middle panels in Figure~\ref{fig:adi} are ages calculated with both $\log g$ and $\rm M_K$ and with $\log g$ only (no $\rm M_K$), respectively. They are both preferring older ages than that of the green isochrone (calculated without $\log g$) which is determined by $\rm M_K$. This demonstrates how including  $\rm M_K$ reduces the effect of potentially unrealistic spectroscopic stellar parameters on age.

The preference for higher ages (when $\log g$ is included in the fitting compared to when it is excluded) seems to decrease at ages greater than 15\,Gyr. We speculate the reason is due to $\rm M_K$ is also forcing higher ages. An example is shown in bottom panels of Figure~\ref{fig:adi}. In addition, the isochrone density increases at higher ages, this means $\rm M_K$ may not be an effective constraint, given parameter uncertainties. 

Over all, this exercise shows that $\rm M_K$ acts as an additional constraint on $\log g$ (and other parameters in general). To further improve the derived ages, we should include even more parameters, such as asteroseimic information and/or multiple $\rm T_{eff}$ and $\log g$ estimates from different methods. \textit{Elli} has been constructed such that adding additional constraints is straightforward. 

\subsection{Age-metallicity relation in the solar vicinity}\label{sec:amr}

As an immediate application of our method, we investigate the age-metallicity relation (AMR) in the solar vicinity, which serves as an important constraint on galactic chemical evolution models. The exact shape of the local AMR in terms of slope and scatter has been extensively debated in literature, with some reporting a rather flat trend for disk stars but with substantial scatter in metallicity across all ages (e.g., C11) while others report a downward trend between age and metallicity, especially at higher ages \citep[e.g.,][]{1993A&A...275..101E,2000A&A...358..850R,2014A&A...565A..89B}. Clearly having reliable stellar ages is crucial in this context as is avoiding potential selection effects. When quantifying the intrinsic cosmic scatter in the AMR it is also important to deconvolve the observational uncertainties (in stellar parameters, including metallicity and age), which is rarely, if ever, done. 

To reconstruct the AMR we rely on the Geneva-Copenhagen Survey \citep{2004A&A...418..989N}, which is ideal for the purpose since it is kinematically unbiased, apparent magnitude limited and volume complete for F and G up to 40\,pc. We use the revised stellar metallicities and effective temperatures of C11 for this dataset and complement them with the improved parallaxes from TGAS, which should lead to more precise stellar ages. Importantly, for the first time we also make allowance for atomic diffusion in the stars \citep{2017arXiv170403465D} by differentiating between the initial bulk metallicity $\rm [Fe/H]_{bulk}$, which is the relevant property for AMR, and the present-day surface metallicity $\rm [Fe/H]_{bulk}$, which is the property measured by observations. Adopting a maximum age uncertainty of 50\% from our Bayesian isochrone fitting limits the C11 sample to 3199 stars; in C11 3236 stars fulfil the same age uncertainty criteria using Hipparcos parallaxes. 

\begin{figure*}
	\includegraphics[width=160mm]{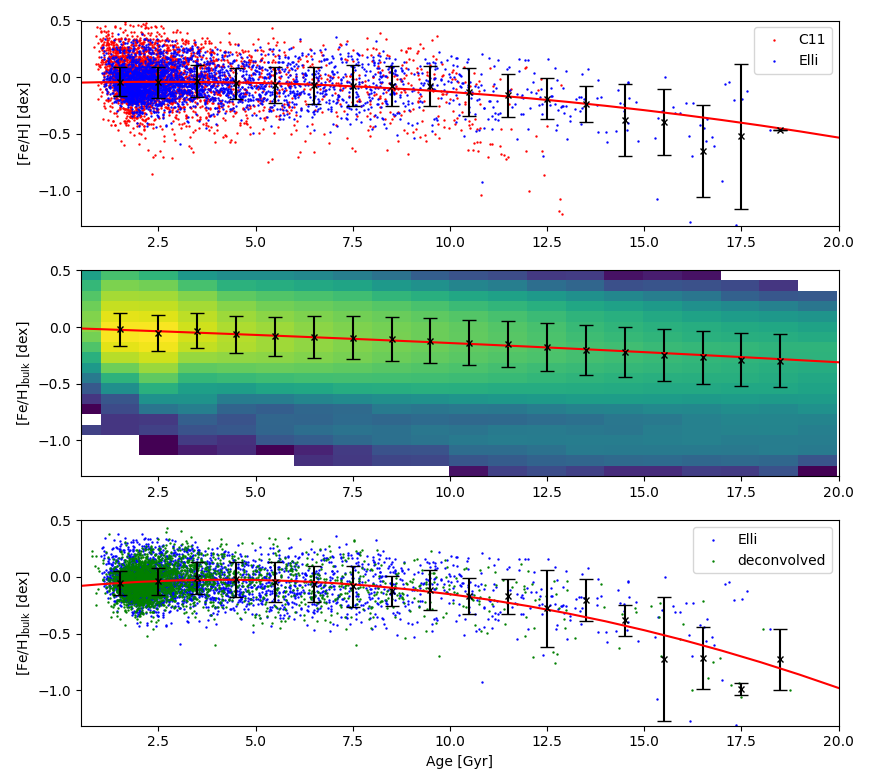}
    \caption{Three different ways of representing the AMR. Black points are 1\,Gyr bin means and standard deviations, red line is the best fitting quadric. \textbf{Top panel:} AMR with \textit{Elli} ages and bulk metallicities (blue) for a set of 4891 stars. AMR from C11 is over plotted in red (only 3236 stars with relative age uncertainty < 50\% are shown), metallicity is taken from C11. \textbf{Middle panel:} Density plot for the \textit{Elli} sample. The densest region is in yellow. \textbf{Bottom panel:} Deconvolved AMR using \textit{Elli} results (green). The original data is in blue. }
    \label{fig:amr}
\end{figure*}

Top panel of Figure~\ref{fig:amr} shows the \textit{Elli}-based AMR as well as the original data from C11; we also plot the mean [Fe/H] and its scatter in 1\.Gyr age bins (black crosses with error bars). Not surprisingly the two analyses agree qualitatively: a nearly flat AMR for ages $<10$\,Gyr with substantial intrinsic scatter at all ages. We applied a quadratic fit to our data and found the fit to be : $\rm [Fe/H]=-0.0016\tau^2  +0.0083\tau-0.0510$ (we have excluded stars with [Fe/H]$<-1.25$ to avoid halo contamination for all fitting purposes). There are however noticeable differences worth discussing. Our ages span a larger range, including to ages $>14$\,Gyr, which were not allowed in C11 due to their age prior. That we obtain such unphysically high ages is largely due to the MIST isochrones that we employ, which ignore alpha-enhancement at low [Fe/H] and assume the low solar abundances of \citet{2009ARA&A..47..481A} while the BASTI isochrones \citep{2004ApJ...612..168P} adopted by C11 are based on the solar abundances of \citet{1998SSRv...85..161G}: for the same stellar parameters MIST isochrones will result in higher ages due to the lower (bulk) metal content than BASTI leading to lower interior opacities. Partly compensating this is the effects of atomic diffusion: as found by \citet{2017arXiv170403465D} properly allowing the surface metallicity to vary along the evolution typically yields lower ages, especially for turn-off stars. Perhaps more noteworthy is that our AMR has smaller [Fe/H] scatter, which we attribute to our consideration of atomic diffusion; we stress that [Fe/H] in Figure~\ref{fig:amr} refers to the initial bulk metallicity in our case while C11 assumes this to be the same as the present-day photospheric [Fe/H] due to their choice of isochrones. Since the amount of scatter and the metal-rich tail of the MDF are supposed to be key signatures of how efficient radial migration is \citep[][C11]{2009MNRAS.399.1145S}, it is clearly important to consider the initial bulk metallicity rather than the present-day surface metallicity, which to our knowledge has not been done previously.

One of the advantageous with a Bayesian approach is the quantification of the uncertainties as probability distribution functions (PDF). In the middle panel of Figure~\ref{fig:amr} we extend the analysis of the AMR by showing it as a density plot making use of the MCMC steps. For each star, 5000 steps in its MCMC chain are plotted instead of its mean age and metallicity to account for the covariances between the two quantities. As before, the mean [Fe/H] and its scatter for each 1\,Gyr age bin is shown (black points) together with a quadratic fit: $\rm [Fe/H]=-0.0002\tau^2  -0.0116\tau-0.0072$. The similarity to the upper panel is attributed primarily to the individual [Fe/H] uncertainties being comparable to or smaller than the intrinsic [Fe/H] scatter, so that the exact shape of the PDF for each datapoint is not critical. 

To truly extract the intrinsic shape of the AMR however requires also to account for the observational errors in age and [Fe/H], which has rarely been attempted before in the literature nor is considered so far in our analysis. In an attempt to do so, here we perform extreme deconvolution \citet{2011AnApS...5.1657B} of the data, by fitting 5 Gaussian components, using the package astroML \citep{astroML}. The covariances of the data points are calculated from their MCMC chains. The resampled, deconvolved data are shown in the bottom panel of  Figure~\ref{fig:amr} together with the mean and scatter for each 1\,Gyr age bin as before. Again, the best fitting quadratic is: $\rm [Fe/H]=-0.0038\tau^2  +0.0325\tau-0.0942$. As expected, the resulting scatter is slightly smaller when accounting for the observational uncertainties while the trend remain largely unaffected. Again, that the scatter is not reduced more stems from the intrinsic [Fe/H] variations dominating over the typical observational errors 
(the [Fe/H] uncertainties from Str\"omgren photometry in C11 are about $\pm 0.1$). 

Finally, selection effects should be considered. The GCS sample is apparent magnitude limited, meaning that metallicity plays a role in the selection function such that given the same age, metal-rich stars are fainter and hence are disproportionally under-represented in the sample. The same effect is confirmed in models for the Gaia-ESO survey by \citet{2014A&A...565A..89B} in which they found the sample completeness at the high age, metal-rich end of the AMR to be less than 50\%. We note that the Gaia-ESO survey has a more complex selection function leading to more pronounced under-sampling but a qualitatively similar bias can be expected in GCS. We therefore conclude that the slopes in the AMR presented in Figure~\ref{fig:amr} are likely slightly over-estimated for high ages. One way to explore the selection effects in a sample like GCS is to perform stellar population synthesis and hierarchical modelling. We intend to carry out such a study with the GALAH survey \citep{2015MNRAS.449.2604D} for the approximately half a million stars contained in the GALAH Data Release 2, roughly two-thirds of which will be dwarfs; besides the enormous sample size, a further advantage with GALAH is the very straightforward selection function. In the meantime, we stress that for any study of the AMR it is important to consider the effects of atomic diffusion and the observational uncertainties as well having accurate stellar parameters, preferably with a Bayesian approach when relying on stellar isochrones. By doing so with the GCS sample, we conclude that while the scatter in metallicity at all ages is substantial, it is less than found by C11. This has implications for how efficient radial migration of stars has been in the Galactic disk.

\section{conclusions and discussions}\label{sec:conc}

This paper describes a new implementation of deriving stellar ages and masses using a Bayesian framework with the code \textit{Elli}. The posterior distributions are sampled using MCMC, which enables us to efficiently calculate full distributions for all parameters involved and quantify uncertainties associated with the values.  Our derived ages and masses for solar neighbourhood stars with spectroscopic parameters and Hipparcos/TGAS parallaxes using  \textit{Elli} are compatible with literature values but with reduced statistical uncertainties in general. Parallax-based $\rm M_K$ seems to be the driving factor behind the deviation between our estimates and literature values. This occurs when $\rm M_K$ and $\log g$ are inconsistent and the parameter with the smaller uncertainty dominates the MCMC algorithm. $\rm M_K$ is further shown to be an additional constraint on parameters, especially when systematic errors are underestimated. 

We present a catalogue of 9111 stars in the solar vicinity with updated ages/masses calculated using both Hipparcos and TGAS parallaxes. We recommend adopting ages computed with TGAS parallaxes, despite them having larger deviations compared to literature values. For the A12 sample in particular, we present only ages determined without $\log g$, since their $\log g$ values are known to be erroneous. The reconstructed AMR of the solar neighbourhood  highlights the flat trend with substantial metallicity scatter for ages $<$ 11\,Gyr. 

The immediate motivation for developing \textit{Elli} has been the imminent arrival of accurate stellar parameters and chemical compositions from large-scale spectroscopic surveys of the Milky Way, in particular the GALAH survey, coupled with the release of exquisite distances using Gaia parallaxes. Such extraordinary data necessitate a corresponding effort in deriving the most accurate possible stellar ages for Galactic archaeology studies. Indeed many of these stars will also have asteroseismic information \citep[e.g.,][]{2016arXiv161003060S,2013ARA&A..51..353C} from the K2 and TESS satellites, which can be straightforwardly incorporated into the analysis with \textit{Elli}, providing complementary age estimations. Future studies will be devoted to the analysis of GALAH stars, including a detailed investigation of the chemical enrichment as a function of time and Galactic location. We particularly stress the great complementarity between the GALAH survey, Gaia and K2, which will be the focus of several future studies in which we will use \textit{Elli} to provide the theoretical foundation for deriving accurate stellar ages for an unprecedented sample.

\section*{Acknowledgements}
We would like to thank Daisuke Kawata for his helpful comments. We acknowledge generous funding from the Australian Research Council through a Laureate Fellowship awarded to MA (grant FL110100012). This work has made use of data from the European Space Agency (ESA) mission {\it Gaia} (\url{http://www.cosmos.esa.int/gaia}), processed by the {\it Gaia} Data Processing and Analysis Consortium (DPAC, \url{http://www.cosmos.esa.int/web/gaia/dpac/consortium}). Funding for the DPAC has been provided by national institutions, in particular the institutions participating in the {\it Gaia} Multilateral Agreement. YST is supported by the Australian Research Council Discovery Program DP160103747, the Carnegie-Princeton Fellowship and the Martin A. and Helen Chooljian Membership from the Institute for Advanced Study at Princeton.

%




\bibliographystyle{mnras}
\bibliography{ref} 




%
%


\bsp	
\label{lastpage}
\end{document}